\documentclass[12pt,twoside,a4paper]{article}

\usepackage[english]{babel}
\usepackage[utf8]{inputenc}
\usepackage{amsmath}
\usepackage{multirow}
\usepackage{tabularx}
\usepackage{colortbl}
\usepackage{amssymb} 
\usepackage{url}
\usepackage[dvips]{graphicx,psfrag}

\usepackage[a4paper, 
            top=4.0cm,
            bottom=3.0cm,
            inner=2.1cm,
            outer=2.1cm,
            bindingoffset=0.77cm]{geometry}

\newenvironment{myquote}%
               {\begin{list}%
                      {}%
                      {%
                       \setlength{\partopsep}{1ex}%
                       \setlength{\topsep}{0.5ex}%
                       \setlength{\parsep}{0.5ex}%
                       \setlength{\leftmargin}{4ex}
                       \setlength{\rightmargin}{4ex}}
                \item}%
               {\end{list}}

\newenvironment{myitemize}%
   {%
    \begin{list}%
      {$\bullet$}%
      {%
       \setlength{\partopsep}{1ex}%
       \setlength{\topsep}{0.2ex}%
       \setlength{\parsep}{0ex}%
       \setlength{\itemsep}{0.5ex}%
       \setlength{\leftmargin}{5ex}%
       \setlength{\itemindent}{0ex}%
       \setlength{\labelsep}{1ex}%
       \setlength{\labelwidth}{5ex}%
      }%
   }{%
     \end{list}%
    }

\newcommand{\captionfonts}{\small}
\makeatletter
\long\def\@makecaption#1#2{%
  \vskip\abovecaptionskip
  \sbox\@tempboxa{{\captionfonts #1: #2}}%
  \ifdim \wd\@tempboxa >\hsize
    {\captionfonts #1: #2\par}
  \else
    \hbox to\hsize{\hfil\box\@tempboxa\hfil}%
  \fi
  \vskip\belowcaptionskip}
\makeatother

\newcommand{\capt}%
           [2]%
           {\centering\parbox{#1}{\caption[dummy]{\small #2}}}

\newcommand{\graycell}{\cellcolor[gray]{.9}}

\newcommand{\etal}{\textit{et al.}}
\newcommand{\OzguvenOY}{Özgüven, Özbak\i{}r, and Yavuz}
\newcommand{\Ozguven}{Özgüven}
\newcommand{\defi}[1]{\textit{#1}}

\newcommand{\conj}[1]{\{ #1 \}}
\newcommand{\abs}[1]{{\mid}#1{\mid}} 
\newcommand{\NN}{N}  
\newcommand{\pp}{p}  
\newcommand{\meanpp}{\bar{\pp}} 
\newcommand{\pprime}{p'} 
\newcommand{\kk}{k} 
\newcommand{\sss}{s} 
\newcommand{\ssr}{s} 
\newcommand{\cc}{c} 
\newcommand{\sstar}{s^*} 
\newcommand{\myel}{l}
\newcommand{\mks}{\mathrm{mks}}
\newcommand{\ff}{f} 
 
\newcommand{\logicand}{\text{ and }}
\newcommand{\Oh}{\mathrm{O}}
\newcommand{\ooy}{ÖÖY}
\newcommand{\ooyprime}{\ooy$'$}

\begin{document}

\begin{center}
{\large \textbf{A MILP model for an extended version \\ of the Flexible Job Shop Problem}\footnote{ Partially funded by CAPES, CNPq, FAPESP, and Hewlett-Packard GOLD Advanced Research grant, through HP Labs and HP Brazil.}}

\vspace*{5mm}
June 28, 2013 
\end{center}
\vspace*{5mm} 
\centerline{Ernesto G. Birgin$^1$, Paulo Feofiloff$^1$, Cristina G. Fernandes$^1$,}
\centerline{Everton L. de Melo$^2$, Marcio T. I. Oshiro$^1$, Débora P. Ronconi$^2$}
\vspace*{2mm}
\centerline{$^1$ Dept.\ of Computer Science, IME, University of São Paulo, Brazil}
\centerline{egbirgin@ime.usp.br, pf@ime.usp.br, cris@ime.usp.br, oshiro@ime.usp.br}
\vspace*{2mm}
\centerline{$^2$ Dept.\ of Production Engineering, EPUSP, University of São Paulo, Brazil}
\centerline{everton.melo@usp.br, dronconi@usp.br}

\begin{abstract}
  A MILP model for an extended version of the Flexible Job Shop Scheduling
  problem is proposed. 
  The extension allows the precedences between operations of a job to
  be given by an arbitrary directed acyclic graph
  rather than a linear order.
  The goal is the minimization of the makespan. Theoretical and practical
  advantages of the proposed model are discussed. Numerical experiments show
  the performance of a commercial exact solver when applied to the proposed
  model. The new model is also compared with a simple extension of the
  model described by \OzguvenOY\ 
  (Mathematical models for job-shop scheduling problems
  with routing and process plan flexibility,
  \emph{Applied Mathematical Modelling}, 34:1539--1548, 2010), 
  using instances from the literature and instances inspired by 
  real data from the printing industry.
\end{abstract}

\section{Introduction}

The Job Shop Scheduling (JS) problem can be stated as follows. 
Consider a set of machines and a set of jobs. 
Each job consists of a sequence of operations to be processed 
in a given order. 
Each operation must be processed individually on a specific machine, 
without preemption. 
The objective is to find a processing sequence for each machine 
that minimizes the makespan, 
which is the completion time of the last operation to be processed. 
The Flexible Job Shop Scheduling (FJS) problem 
is a generalization of the JS problem 
in which there may be several machines, not necessarily identical, 
capable of processing an operation. 
Specifically, for each operation, 
we are given the set of machines on which that operation can be processed.
The goal is to decide on which machine each operation will be processed, 
and in what order the operations will be processed on each machine, 
so that the makespan is minimized.

The JS problem is known to be NP-hard~\cite{garey:76}.
It is one of the most difficult combinatorial optimization problems 
according to Lawler \etal~\cite{lawler:93}. 
Since the FJS problem is at least as difficult as the JS, 
it is also NP-hard. 
Many researchers use heuristic methods 
to deal with (the minimization of makespan of) the FJS problem.
See for example
\cite{brandimarte:93,ChanWC06,GutirrezGM11,MishraP89,VilcotB08,ZhangSLG}. 
In contrast, the number of publications 
concerned with the exact solution of the FJS problem is very small. 
Fattahi, Mehrabad, and Jolai~\cite{fattahi:07}
presented one of the most relevant papers in this direction. 
They proposed a mixed integer linear programming (MILP) model 
for the FJS problem and used it to solve 
a set of 20 instances of small and medium size with the LINGO software. 
The results were compared to those obtained by heuristic methods.

\OzguvenOY~\cite{ozguven:10} 
elaborated a more concise MILP model for the FJS problem, 
modifying an earlier one by Manne~\cite{manne:60} 
to incorporate machine flexibility. 
We shall name their model~\ooy. 
\Ozguven~\etal\ tested their model by solving the 20 instances 
mentioned above with the CPLEX solver.
They obtained optimal solutions for the 10 small size instances 
faster than Fattahi~\etal\
They also obtained optimal solutions for five of the medium size instances 
and better bounds for the other five, 
while Fattahi \etal\ 
did not find optimal solutions for any of those instances.

\begin{figure}[tb]
\begin{center}
\begin{tabular}{ccc}
\includegraphics[scale=0.7]{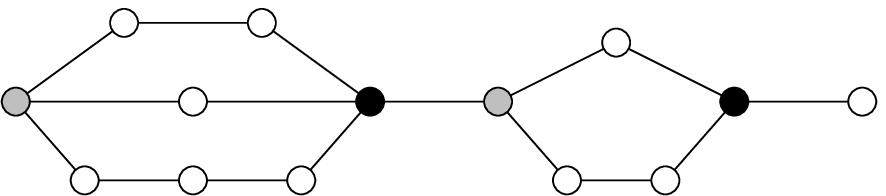} && 
\includegraphics[scale=0.7]{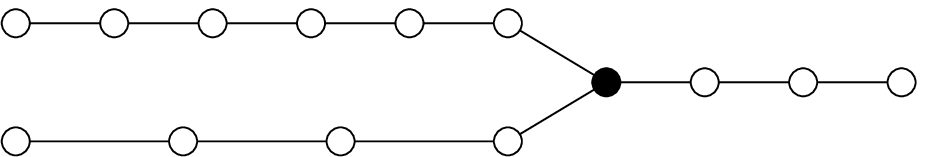} \\
\small{(a)}  & & \small{(b)}  \\
\end{tabular}
\caption{\label{forkjob} (a) A representation of a more general type
  of job.  Each node represents an operation.  The arcs represent
  precedence constraints and all arcs are directed from left to right.
  The black nodes are assembling operations and the gray nodes are
  disassembling operations.  (b)~A representation of a Y-job.}
\end{center}
\end{figure}

In the literature, each job in the FJS problem consists of 
a sequence of operations to be processed 
in a given order, just as in the ordinary JS problem.
In an industrial environment, however,
it is common to have jobs some of whose operations 
can be processed simultaneously. 
Moreover, some mutually independent sequences of operations
may feed into an ``assembling'' operation.
Similarly, there may be ``disassembling'' operations which split 
the sequence of subsequent operations into 
two or more mutually independent sequences. 
A representation of this kind of job is shown in Figure~\ref{forkjob}(a). 

In a real problem from the printing industry~\cite{ZengJLGHM10},
for example, some jobs consist of two 
independent sequences of operations
followed by a third that puts together the results of the first two. 
A representation of this kind of job 
is shown in Figure~\ref{forkjob}(b). 
We say that such configuration is a \defi{Y-job},
while the traditional type of job (a simple sequence of operations) 
is a \defi{path-job}.  

Vilcot and Billaut~\cite{VilcotB08} have already considered a class of instances 
that includes Y-jobs and path-jobs. 
They describe an environment, 
coming from the printing and boarding industry, 
where each operation in a job can have more than one predecessor, 
but at most one direct successor. 
Alvarez-Valdés \etal~\cite{Alvarez-ValdesFTGR05} describe yet another environment, 
coming from a glass factory, 
that requires an even more general variant of the FJS problem.

In this paper, we extend the usual definition of the FJS problem to allow a
job to be a set of operations with an arbitrary precedence relation, thus
including general types of jobs like those in Figure~\ref{forkjob}. 
Then we propose a new MILP model for the extended FJS problem that focuses on
the operations and the precedences among them, 
and leaves the jobs only implicitly defined.

Difficult problems such as JS and FJS 
require sophisticated techniques to produce 
reasonable results in acceptable time. 
It is usually preferable to get good sub-optimal 
results quickly than to wait for days to get an optimal solution.  
However, to evaluate the quality of the non-exact methods, 
it is desirable to compare their results with exact solutions 
(at least for some reference set of small and medium size instances).
This is one of the main reasons for developing exact
methods for the FJS problem. Good mixed integer programming models, combined
with the availability of powerful MILP solvers, are capable of providing, 
if not an optimal solution, at least good bounds to many small and 
medium size instances. 
These bounds can then be used to evaluate the quality of heuristic methods.

The goal of this paper is to contribute to the development of exact models for
the FJS problem, and to present a new set of instances to be used in 
comparisons between different MILP models and heuristics for the problem.
Section~\ref{sec:notation} introduces some notation, 
while Section~\ref{sec:model} presents the new model 
we propose for the problem.
Section~\ref{sec:model} also describes an adaptation of the \ooy\  model 
needed to address the more general types of jobs. 
Section~\ref{sec:exper} presents the results of a computational experiment
involving the two models, using as benchmark 
the 20 instances of Fattahi \etal, 
the 15 instances of Brandimarte~\cite{brandimarte:93}, 
and a new set of 50 instances that contain jobs 
such as the ones depicted in Figure~\ref{forkjob}.

\section{Notation} 
\label{sec:notation}

Let $(V,A)$ be a dag, i.e.\ a directed acyclic graph. 
The vertices of the dag represent the \defi{operations}. The arcs represent 
\defi{precedence constraints}. 
(One can think of a \defi{job} as a weakly connected component
of the dag, but this concept is not needed for the models.)  We are
also given a set $M$ of \defi{machines} and a function $F$ that
associates a non-empty subset~$F(v)$ of $M$ with each operation~$v$. 
The machines in $F(v)$ are the ones that can process operation~$v$. 
Additionally, for each operation $v$ and each machine $\kk$ in~$F(v)$, 
we are given a positive rational number~$\pp_{v,\kk}$
representing the \defi{processing time} of operation $v$ on machine~$\kk$.

A \defi{machine assignment} is a function $\ff$ 
that assigns a machine $\ff(v)\in F(v)$ with 
each operation~$v$.
Given a machine assignment~$\ff$, let $\pp^{\ff}_v := \pp_{v,\ff(v)}$.

For each machine $\kk$, 
let $V_{\kk}$ be the set of operations that can be processed on 
machine~$\kk$, that is, $V_{\kk} = \conj{v\in V : \kk \in F(v)}$.
Let $B_{\kk}$ be the set of all ordered pairs of distinct elements of~$V_{\kk}$. 
The pairs $(v,w)$ in $B_{\kk}$ are designed to prevent $v$ and $w$ 
from using machine $\kk$ at the same time.  
Let $B$ denote the union of all~$B_{\kk}$.  
Hence, $(v,w)\in B$ if and only if $v \neq w$ 
and $F(v)\cap F(w) \neq \emptyset$.

Given a machine assignment~$\ff$, let $B^{\ff}$ be the set of all ordered
pairs of distinct operations to be processed on the same machine, 
that is, $B^{\ff} = \conj{(v,w)\in B : \ff(v)=\ff(w)}$. 
A \defi{selection} is any subset $Y$ of $B^{\ff}$ such that,
for each $(v,w)$ in $B^{\ff}$, exactly one of $(v,w)$ and $(w,v)$ is in~$Y$. 
A selection corresponds to an ordering of the operations 
to be processed on the same machine. 
A selection $Y$ is \defi{admissible} if $(V, A\cup Y)$ is a dag. 
In other words, a selection is admissible if 
it conflicts neither with the given precedence constraints nor with itself. 

Given a machine assignment $\ff$ and an admissible selection $Y$, 
a \defi{schedule} for $(V, A \cup Y, \pp^{\ff})$ 
is a function~$\sss$ from $V$ to the set of
nonnegative rational numbers 
such that $\sss_v + \pp^{\ff}_v \leq \sss_w$ for each $(v,w)$ in $A \cup Y$.  
(A~schedule exists since $(V,A \cup Y)$ is a dag.)  
The number $\sss_v$ is the \defi{starting time} of operation~$v$.
(We write ``$\sss_v$'' instead of the more traditional~``$\sss(v)$''.)
The \defi{makespan} of a
schedule~$\sss$ is the number $\mks(\sss) := \max_{v\in V}\:(\sss_v + \pp^{\ff}_v)$.
This definition does not preclude 
idle time in the schedule;
the next one focuses on nondelay schedules. 

The \defi{length} of a (directed) path
$(v_1,v_2,\ldots,v_\myel,v_{\myel+1})$ in the dag $(V,A \cup Y)$ is the
number $\pp^{\ff}_{v_1}+\pp^{\ff}_{v_2}+\cdots+\pp^{\ff}_{v_\myel}$.  
(Note that $\pp^{\ff}_{v_{\myel+1}}$ is not part of the sum.)  For any path $P$
in~$(V,A \cup Y)$ ending at $v$ and any schedule~$\sss$, the length of
$P$ is at most~$\sss_v$.
For each $v$ in $V$, let $\sstar_v$ be the maximum of the lengths of all
paths in $(V, A \cup Y)$ ending at $v$.  The function $\sstar$ so
defined is a schedule.  We say that this is the \defi{tight schedule}
for~$(V,A \cup Y,\pp^{\ff})$.  There is a simple dynamic programming
algorithm that computes the tight schedule.  
Not surprisingly, the makespan of the tight schedule $\sstar$
is determined by long paths: there exists a path
$P=(v_1,v_2,\ldots,v_\myel,v_{\myel+1})$ in $(V, A \cup Y)$ such that
the length of $P$ plus $\pp_{v_{\myel+1}}$ equals $\mks(\sstar)$.
(Such $P$ is known as a \defi{critical} path.)  
It follows from the previous observations that the tight schedule 
has minimum makespan among all schedules for~$(V, A \cup Y, \pp^{\ff})$.

The \defi{makespan} of an admissible selection $Y$ for a given machine
assignment $\ff$, denoted by $\mks(Y)$, is the makespan of the tight
schedule for $(V, A\cup Y, \pp^{\ff})$. The FJS problem can now be stated
as follows:

\begin{myquote}
\textsc{FJS Problem}  $(V, A, M, F, \pp)$:
Find a machine assignment  $\ff$
and an admissible selection~$Y$ 
such that $\mks(Y)$ is minimum.
\end{myquote}

\section{Models}\label{sec:model}

This section presents two models for the FJS problem. The first one is our new
model while the second is an adapted version of the \ooy\  model by \OzguvenOY.

\subsection{A new model for the FJS problem}
\label{sec:our-f-jssp-milp}

In order to formulate the FJS problem as a MILP, we use a binary array $x$
to represent the machine assignments and a binary array $y$ to represent
selections.  The first array will have a component $x_{v,\kk}$ for each $v$
in $V$ and each $\kk$ in~$F(v)$.  The second will have a component $y_{v,w}$
for each $(v,w)$ in~$B$.  
We also use two rational arrays $\sss$ and $\pprime$,
and a rational number~$z$, where 
$\sss$ represents the starting times, 
$\pprime$ represents the processing times 
corresponding to the machine assignment given by~$x$, 
and $z$ represents the makespan of schedule~$\sss$.

We need an upper bound $L$ on the makespan of an optimal solution of
the FJS problem.  This can be the makespan of an arbitrary admissible
selection or, alternatively, a global bound like $\sum_{v\in V}
\max_{\kk\in F(v)}\pp_{v,\kk}$.

Our MILP can now be given as follows: find a rational number $z$, rational
arrays $\sss$ and $\pprime$, and binary arrays $x$ and $y$ that
\begin{eqnarray}
\mbox{minimize $z$} \hspace{4cm} && \nonumber \\
\mbox{subject to} \hspace{4cm} && \nonumber \\
s_v + \pprime_v \leq z
  && \mbox{$\forall\, v \in V$,} \label{eq3a:objective}\\ 
\textstyle\sum_{\kk\in F(v)} x_{v,\kk}= 1
  && \mbox{$\forall\, v \in V$,} \label{eq3a:mach-injection}\\
\pprime_v = \textstyle\sum_{\kk\in F(v)} \pp_{v,\kk}x_{v,\kk}
  && \mbox{$\forall\, v \in V$,} \label{eq3a:effective-cost}\\
y_{v,w} + y_{w,v} \geq x_{v,\kk}+x_{w,\kk} - 1
  && \mbox{$\forall\, \kk \in M$ and $\forall\, (v,w) \in B_{\kk}$,} \label{eq3a:x-plus-x-geq-1}\\
s_v + \pprime_v \leq s_w
  && \mbox{$\forall\, (v,w) \in A$,} \label{eq3a:hard-precedence}\\ 
s_v + \pprime_v - (1-y_{v,w})L \leq s_w
  && \mbox{$\forall\, (v,w) \in B$,} \label{eq3a:computed-precedence}\\
s_v \geq 0
  && \mbox{$\forall\, v \in V$.} \label{eq3a:nonneg-start}
\end{eqnarray}

As $x$ is binary, constraint~(\ref{eq3a:mach-injection}) 
ensures that $x$ is a machine assignment. 
Then constraint~(\ref{eq3a:effective-cost}) 
makes array $\pprime$ represent the processing times of operations. 
In fact, $\pprime$ can be seen as an intermediate value,
not a variable, that helps to simplify the presentation of the model. 
Since $\pp_{v,\kk} > 0$ for all $v$ and~$\kk$, thus $\pprime_v > 0$ and so 
constraint~(\ref{eq3a:computed-precedence})
makes sure that $y_{v,w}$ and $y_{w,v}$ are not both equal to~$1$. 
Hence, as $y$ is binary, 
constraint~(\ref{eq3a:x-plus-x-geq-1}) implies that $y$ represents a selection. 
Indeed, if $x_{v,\kk} = x_{w,\kk} = 1$, 
which means $v$ and $w$ are assigned to machine~$\kk$, 
then~(\ref{eq3a:x-plus-x-geq-1}) forces $y$ to decide 
whether $v$ comes before or after~$w$. 
Otherwise (i.e.\ if $x_{v,\kk}$ and $x_{w,\kk}$ are not both~$1$),
constraint~(\ref{eq3a:x-plus-x-geq-1}) is trivially satisfied.
Once $y$ is a selection and~$\pprime$ represents the processing times, 
constraints~(\ref{eq3a:hard-precedence}), (\ref{eq3a:computed-precedence}), 
and~(\ref{eq3a:nonneg-start}) make $\sss$ represent a schedule. 
Finally, the objective function and constraint~(\ref{eq3a:objective}) 
make sure $z$ is the makespan of the schedule, and is
as small as possible.

We show next that our MILP is equivalent to the~FJS problem.
Suppose $(\ff,Y)$ is a feasible solution of 
an instance $(V, A, M, F, \pp)$ of the problem,
i.e.\ suppose $\ff$ is a machine assignment 
and $Y$ a corresponding admissible selection.
Let $\sss$ be the tight schedule for $(V,A\cup Y,\pp^{\ff})$
and let $z := \mks(\sss)$.
For each $v$ and each $\kk$ in $F(v)$, 
let $x_{v,\kk}:=1$ if and only if $\ff(v)=\kk$.  
Let $y_{v,w}:=1$ for each $(v,w)$ in $Y$ 
and $y_{v,w}:=0$ for each $(v,w)$ in~$B \setminus Y$.
Finally, let $\pprime := \pp^{\ff}$. 
Then, the tuple $(\sss, x, \pprime, y, z)$ 
is a feasible solution of our MILP 
and the value, $z$, of this solution
is equal to~$\mks(\sss)$.
\
Now consider the converse. 
Let $(\sss, x, \pprime, y, z)$ be a feasible solution of the~MILP.  
For each~$v$, let $\ff(v)$ be the unique~$\kk$ in $F(v)$ for which $x_{v,\kk}=1$; 
such $\kk$ exists because $x$ is binary 
and constraint~(\ref{eq3a:mach-injection}) holds.  
According to~(\ref{eq3a:effective-cost}), $\pprime = \pp^{\ff}$.  
Let $Y$ be the set of all pairs $(v,w)$ in $B$ such that $y_{v,w}=1$.
Constraint~(\ref{eq3a:x-plus-x-geq-1}) makes sure that $Y$ is a selection.
Since $\pp_{v,\kk}>0$ for all~$(v,\kk)$,
constraints~(\ref{eq3a:hard-precedence}) to~(\ref{eq3a:nonneg-start}) make
sure that the selection $Y$ is admissible.
Hence, $(\ff,Y)$ is a feasible solution of the~FJS problem.
Moreover, $\sss$ is a schedule for~$(V, A\cup Y, \pprime)$
and $\mks(\sss) \leq z$ by virtue of~(\ref{eq3a:objective}).
\
This discussion shows that
every optimal solution of the FJS problem 
corresponds to an optimal solution of the MILP
and conversely.

Let $\beta := \sum_{\kk \in M}\abs{B_{\kk}}$ 
and $\varphi := \sum_{v \in V}\abs{F(v)}$.  
Our MILP has $2\abs{V} + \abs{A} + \abs{B} + \beta$ constraints, and $\abs{V}
+ \varphi + \abs{B}$ variables, of which $\varphi + \abs{B}$ are binary.

\subsection{The model of \OzguvenOY}
\label{sec:adapted}

The \ooy\ MILP model mentioned in the Introduction is designed to
handle path-jobs only.  In order to
represent 
our more general job
structure, a slight modification of that MILP is needed. (Thus, for
example, the double 
index 
$ij$, representing operation $j$ of
job~$i$, 
was 
replaced by a single index $v$ representing an
operation.  Accordingly, the set $O_i$ of 
operations 
in job $i$
was replaced by
our set $A$ of precedence constraints combined with the set
$B_k$. Finally, the role of the ``big number'' $L$ was made precise.)
We will refer to the modified version of the \ooy\ model
as~\ooyprime. We introduce~\ooyprime\ only 
because we are unable 
to compare our
proposed model with \ooy\ 
directly. 
The modified model reduces to the original one 
when applied to instances with path-jobs only.

We shall use the following variables. 
First, a binary array $x$ to represent machine assignments, 
with one component $x_{v,\kk}$ for each $v$ in $V$ and each $\kk$ in~$F(v)$. 
Second, a rational array $\ssr$ to represent the starting times, 
with one component $\ssr_{v,\kk}$ for each $v$ in $V$ and each $\kk$ in~$F(v)$.
Third, a rational array $t$ to represent the completion times, with
one component $t_{v,\kk}$ for each $v$ in $V$ and each $\kk$ in~$F(v)$. 
Finally, a binary array $y$ to represent a selection,
with one component $y_{v,w,\kk}$ for each $\kk$ in $M$ and $(v,w)$ in $B_{\kk}$. 

We also need an upper bound $L$ on the makespan of an optimal solution.  
Again, this can be the makespan of some admissible selection or
the sum $\sum_{v\in V} \max_{\kk\in F(v)}\pp_{v,\kk}$, for example.
Finally, let $\hat{V}$ be the set of ``terminal'' vertices, 
i.e.\ the set of all $v$ in $V$ such that there is no $(v,u)$ in~$A$.

The \ooyprime\ model can be stated as follows: find a rational number $z$,
rational arrays $\ssr$ and $t$ and binary arrays $x$ and~$y$ that
\begin{eqnarray}
\mbox{minimize $z$} \hspace{4cm} && \nonumber \\
\mbox{subject to} \hspace{4cm} && \nonumber \\
t_{v,\kk} \leq z
  && \mbox{$\forall\, v \in \hat{V}$ and $\forall\, \kk \in F(v)$,} \label{eq4:objective}\\ 
\textstyle\sum_{\kk\in F(v)} x_{v,\kk}= 1
  && \mbox{$\forall\, v \in V$,} \label{eq4:mach-injection}\\
\ssr_{v,\kk} + t_{v,\kk} \leq 2x_{v,\kk}L
  && \mbox{$\forall\, v \in V$ and $\forall\, \kk \in F(v)$,} \label{eq4:wrong-machine-s}\\
y_{v,w,\kk} + y_{w,v,\kk} = 1
  && \mbox{$\forall\, \kk \in M$ and $\forall\, (v,w) \in B_{\kk}$,} \label{eq4:y-plus-y-is-one}\\
\ssr_{v,\kk} + \pp_{v,\kk} - (1-x_{v,\kk})L \leq t_{v,\kk}
  && \mbox{$\forall\, v \in V$ and $\forall\, \kk \in F(v)$,} \label{eq4:s-of-v-leq-t-of-v}\\
t_{v,\kk} - (1-y_{v,w,\kk})L \leq \ssr_{w,\kk}
  && \mbox{$\forall\, \kk \in M$ and $\forall\, (v,w) \in B_{\kk}$,} \label{eq4:soft2}\\
\textstyle\sum_{\kk\in F(v)} t_{v,\kk} \leq \sum_{\kk\in F(w)} \ssr_{w,\kk} 
  && \mbox{$\forall\, (v,w) \in A$,} \label{eq4:hard-precedence}\\
\ssr_{v,\kk} \geq 0
  && \mbox{$\forall\, v \in V$ and $\forall\, \kk \in F(v)$,} \label{eq4:nonneg-start}\\ 
t_{v,\kk} \geq 0
  && \mbox{$\forall\, v \in V$ and $\forall\, \kk \in F(v)$.} \label{eq4:nonneg-completion}
\end{eqnarray}

As $x$ is binary, constraint~(\ref{eq4:mach-injection}) ensures that $x$ is a
machine assignment. If $v$ is not assigned to machine $\kk$, 
constraint~(\ref{eq4:wrong-machine-s}) makes $\ssr_{v,\kk}=t_{v,\kk}=0$. 
Thus, at most one term is nonzero in each sum of~(\ref{eq4:hard-precedence}), 
and constraint~(\ref{eq4:wrong-machine-s}) is trivial 
if $\ssr_{v,\kk}=t_{v,\kk}=0$.  
Given a machine assignment $x$, the set of pairs $(v,w)$ 
such that $y_{v,w,\kk}=1$ and both $v$ and $w$ are assigned to machine~$\kk$
is a selection due to constraint~(\ref{eq4:y-plus-y-is-one}).
Now, constraints~(\ref{eq4:s-of-v-leq-t-of-v}) to~(\ref{eq4:nonneg-completion})
make $\ssr$ store the starting times and $t$ the completion times of a
schedule.  Indeed, (\ref{eq4:s-of-v-leq-t-of-v}) ensures that the starting and
completion times of an operation are consistent, (\ref{eq4:soft2}) ensures
that two operations assigned to the same machine have starting and completion
times consistent with the selection, and~(\ref{eq4:hard-precedence}) ensures
that the starting and completion times satisfy the precedences between
operations. Finally, constraint~(\ref{eq4:objective}) and the objective
function ensure that $z$ is the makespan of the schedule, and is as small as
possible.

We show next that the \ooyprime\ MILP is equivalent to the FJS problem.
Suppose $(\ff,Y)$ is a feasible solution of an instance $(V, A, M, F, \pp)$ 
of the problem.
Let $\sstar$ be the tight schedule for~$(V,A\cup Y, \pp^{\ff})$
and $z := \mks(\sstar)$.
For each~$v$ and each $\kk$ in~$F(v)$, 
let $x_{v,\kk}:=1$ if and only if $\ff(v)=\kk$.  
For each~$v$ and each~$\kk$ in~$F(v)$, 
let $\ssr_{v,\kk}:=\sstar_v$ if $\ff(v)=\kk$ 
and $\ssr_{v,\kk}:=0$ otherwise.  
Similarly, let $t_{v,\kk}:=\sstar_v+\pp_{v,\kk}$ if $\ff(v)=\kk$ 
and $t_{v,\kk}:=0$ otherwise.  
Then $\ssr$, $t$, $x$, and $z$ satisfy 
constraints~(\ref{eq4:objective}) to~(\ref{eq4:wrong-machine-s}),
(\ref{eq4:s-of-v-leq-t-of-v}),
and~(\ref{eq4:hard-precedence}) to~(\ref{eq4:nonneg-completion}).  
Now define $y$ as follows.  
Let $v_1,\, v_2,\, \ldots,\, v_{\NN}$ be an arbitrary ordering 
of all the operations in~$V$.  
For each $\kk$ in~$M$ and each $(v_i,v_j)$ in~$B_{\kk}$, 
let $y_{v_i, v_j, \kk} := 1$ if and only if one of
the following three conditions holds: 
$(v_i, v_j) \in Y$ and $\ff(v_i) = \ff(v_j) = \kk$; 
or $\ff(v_i) \neq \kk$ and $\ff(v_j) = \kk$; 
or $i > j$, $\ff(v_i) \neq \kk$ and $\ff(v_j) \neq \kk$.
This definition of $y$ satisfies
constraints~(\ref{eq4:y-plus-y-is-one}) and~(\ref{eq4:soft2}).
Hence, the tuple $(\ssr, t, x, y, z)$ 
is a feasible solution of the MILP
and the value, $z$, of this solution
is equal to~$\mks(\sstar)$.
\
Now consider the converse. 
Let $(\ssr, t, x, y, z)$ be a feasible solution of the MILP.
For each~$v$,
let $\ff(v)$ be the unique $\kk$ in $F(v)$ for which $x_{v,\kk}=1$;
such $\kk$ exists because $x$ is binary and~(\ref{eq4:mach-injection}) holds.
For each~$\kk$,
let $Y_{\kk}$ be the set of all pairs $(v,w)$ in~$B_{\kk}$
such that $x_{v,\kk}=1$, $x_{w,\kk}=1$, and $y_{v,w,\kk}=1$.
Let $Y$ be the union of all $Y_{\kk}$, $\kk\in M$.
According to~(\ref{eq4:y-plus-y-is-one}), $Y$ is a selection.
According to~(\ref{eq4:s-of-v-leq-t-of-v}), (\ref{eq4:soft2}),
and~(\ref{eq4:hard-precedence}),
$Y$ is an admissible selection.
Hence, $(\ff,Y)$ is a feasible solution of the~FJS problem.
Finally, let $\ssr'_v := \ssr_{v,\ff(v)}$
and observe that $\ssr'$ is a schedule for~$(V, A\cup Y, \pp^{\ff})$
and $\mks(\ssr') \leq z$.
\
This discussion shows that
every optimal solution of the FJS problem 
corresponds to an optimal solution of the MILP
and conversely.

As before, let $\beta := \sum_{\kk \in M}\abs{B_{\kk}}$ and $\varphi := \sum_{v \in V}
\abs{F(v)}$. Additionally, let $\hat{\varphi} := \sum_{v \in \hat{V}} \abs{F(v)}$. 
This MILP has $\abs{V} + \abs{A} + \hat{\varphi} + 2\varphi + 2\beta$
constraints, and $3\varphi+\beta$ variables, 
of which $\varphi+\beta$ are binary.

\subsection{A brief comparison of the two models}

In the next section, we compare the two models experimentally. Before that, 
we comment on some differences between the models.

The proposed model, presented in Section~\ref{sec:our-f-jssp-milp}, is more
compact than the \ooyprime\ model. The array variables that describe the
selection and the starting times have one more dimension in \ooyprime,
since they are also indexed by the machines. 
Moreover, \ooyprime\ has variables for the completion times
that depend not only on the operations but also on the machines.
Therefore, the \ooyprime\ model uses significantly more variables. Concretely,
since $\varphi \geq \abs{V}$ and $\beta \geq \abs{B}$, the \ooyprime\ model
uses at least $\varphi$ more variables, at least as many binary variables, and
has at least $\hat{\varphi} + \varphi$ more constraints than our model.
 
Another difference between the two models comes from the following
observations. Consider the linear relaxation of the proposed MILP, 
i.e.\ drop the requirement that $x$ and~$y$ be integer.  
Let $(\dot{\sss}, \dot{x}, \dot{\pp}', \dot{y}, \dot{z})$ 
be an optimal solution of the resulting~LP.
As~(\ref{eq3a:hard-precedence}) enforces the precedence constraints in~$A$, 
considering the processing times given by~$\dot{\pp}'$, 
the starting times given by $\dot{\sss}$ are a schedule for $(V,A,\dot{\pp}')$. 
Moreover, 
by~(\ref{eq3a:objective}), $\mks(\dot{\sss}) \leq \dot{z}$.
Now let $\sstar$ be the tight schedule for $(V,A,\dot{\pp}')$.  
Then of course $\mks(\sstar) \leq \mks(\dot{\sss}) \leq \dot{z}$.  
Hence, the optimal value, $\dot{z}$, 
of the relaxed MILP is not smaller than 
the makespan of the tight schedule for $(V,A,\dot{\pp}')$.  
(In fact, $\dot{z}$ is also not smaller than the makespan of 
the tight schedule for $(V,A,\pp'')$, 
where $\pp''_v = \min_{\kk\in F(v)}\pp_{v,\kk}$.)

Similarly, consider the linear relaxation of the \ooyprime\ MILP, 
i.e.\ drop the requirement that $x$ and $y$ be integer.  
Now take any instance of the FJS problem in which
\begin{myitemize}
\item
$\pp_{v,\kk} \leq L/2$ for each $v$ and each $\kk$ and
\item
$\abs{V_{\kk}} \geq 2$ for each $\kk$ in $M$, 
\end{myitemize}
where $V_{\kk}:=\conj{v\in V: \kk \in F(v)}$.
(We believe many instances of the FJS problem
do possess these properties.)
Then the linear relaxation of the \ooyprime\ MILP
has an optimal solution $(\dot{\ssr}, \dot{t}, \dot{x}, \dot{y}, \dot{z})$ 
with $\dot{\ssr}=0$, $\dot{t}=0$, 
$\dot{x}_{v,\kk}=1/\abs{V_{\kk}}$, $\dot{y}_{v,w,\kk}=1/2$, and $\dot{z} = 0$. 
That is, for these instances, 
the optimal value of the relaxation of the \ooyprime\ MILP 
is potentially much smaller than the optimal value of the relaxation 
of the proposed MILP. 
Moreover, the integrality gap of the \ooyprime\ MILP is unbounded. 
This seems to be an undesirable property of this MILP, 
a property not shared by the proposed MILP.

\section{Computational experiments}\label{sec:exper}

This section presents the results of a computational experiment involving 
the two MILP models. 
We used as benchmarks the 20 instances of Fattahi \etal~\cite{fattahi:07}, 
the 15 instances of Brandimarte~\cite{brandimarte:93}, 
and a new set of 50 instances that contain jobs 
as the ones depicted in Figure~\ref{forkjob}. 

In an application of the FJS problem coming from the printing
industry~\cite{ZengJLGHM10}, each job represents an order for a certain number
of printed copies of some object.  
The way to process each order depends on the type of object to be printed.  
Some of the orders correspond to a path-job.
Some others, like the printing of a book,
correspond to a Y-job as in Figure~\ref{forkjob}(b) with the following parts:
the printing of the book cover, 
the printing of the text blocks (book pages),
and the binding. 
There is more than one choice of machine to process some of the operations.

To analyze the performance of the MILP models on this kind of
instances, we developed a generator of random instances of Y-jobs.  
The generator has four integers as parameters: 
the number $n$ of jobs, 
the number $o$ of operations per job, 
the number $m$ of machines, 
and the maximum number $q$ of machines that can process the same operation.  
First, the dag representing the jobs is created. 
Each job is initially a path-job with operations $1$, $2$, \nolinebreak\ldots,~$o$.
Then, for each job,
two operations $i$ and $j$ are chosen independently at random.
If $i=1$, $j=1$, or $i=j$, the job is not changed.
Else, assuming $i< j$,
arc $(i{-}1,i)$ is replaced by arc $(i{-}1,j)$.
Next, for each operation, 
a set of at most $q$ machines is selected at random
and for each of these machines a processing time is chosen 
at random in the set $\conj{20, \ldots, 200}$. 
(All random choices use a uniform distribution.)
A set of $20$ instances of this type, 
with $n$ and $o$ in $\conj{4,\ldots,17}$,
$m$ in $\conj{7,\ldots,26}$,
and $q$ in $\conj{3,5,8}$,
was generated.
The instances were named YJS01 through YFJS20.

\begin{figure}[t]
\begin{center}
\begin{tabular}{ccccc}
\includegraphics[scale=0.8]{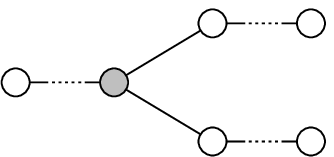} && 
\includegraphics[scale=0.8]{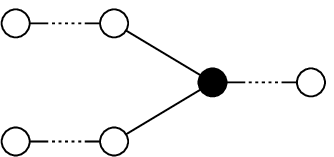} &&
\includegraphics[scale=0.8]{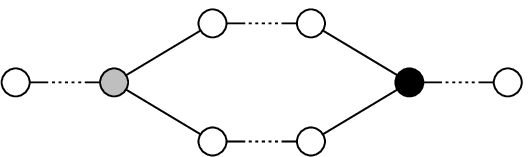} \\
\footnotesize D2 && \footnotesize A2 && \footnotesize DA2 \\[2mm]
\includegraphics[scale=0.8]{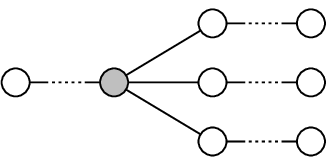} && 
\includegraphics[scale=0.8]{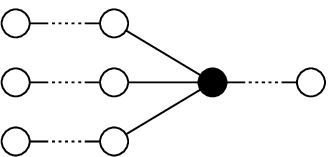} && 
\includegraphics[scale=0.8]{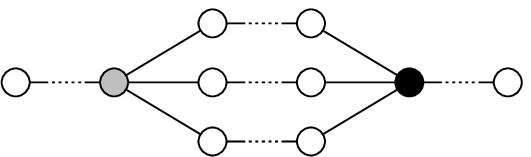} \\
\footnotesize D3 && \footnotesize A3 && \footnotesize DA3 \\
\end{tabular}
\caption{\label{second-generator} The six types of dags~--- D2, D3,
  A2, A3, DA2, and DA3~--- produced by our second generator of
  instances. Each node represents an operation. The arcs represent
  precedence constraints and are all directed from left to right. In
  each dag, all maximal paths have the same length. (The black nodes
  are assembling operations and the gray nodes are disassembling
  operations.)}
\end{center}
\end{figure}

We also developed a generator of random instances of a more general kind.
The generator has two integer parameters:
the number $n$ of jobs and the number $m$ of machines. 
Each job is represented by a dag of one of the six possible types~---
D2, D3, A2, A3, DA2, and~DA3~---
indicated in Figure~\ref{second-generator}.
In each dag,
all maximal paths have the same length.
(Hence, in DA3, for example, 
the three ``parallel'' paths in the middle section
have the same number of operations.)
For each job, one of the six types of dag is chosen at random.
Then, the length of the maximal paths through the dag is chosen
at random from the set $\conj{\lceil m/2 \rceil, \ldots, m}$.
The lengths of the sections of the dag
(left and middle sections 
in case of DA and left in case of D and~A)
are also chosen at random.
For each operation~$v$,
the size of the set $F(v)$ is chosen at random in 
$\conj{\lceil 0.3m \rceil, \ldots, \lceil 0.7m \rceil}$
and then the elements of $F(v)$ are chosen at random 
from $\conj{1,2,\ldots,m}$.
The processing time of $v$ on one of the machines in $F(v)$
is set to a random number~$p$ 
in $\conj{1,\ldots,99}$. 
The processing times of $v$ on 
each of the remaining machines in~$F(v)$
is a random number 
in $\conj{p,\ldots,\min(3p, 99)}$. 
(All random choices use uniform distribution.)
A set of~$30$ instances was generated in the manner just described, 
with~$n$ in~$\conj{4,\ldots, 12}$ 
and~$m$ in~$\conj{5,\ldots, 10}$. 
The instances were named DAFJS01 through DAFJS30
and used in the computational experiments presented next.

%
%
%

\medskip

To solve the MILPs, we used the IBM ILOG CPLEX 12.1 solver with the
following settings: 3600s for time limit, 1 for maximum number of
threads, and 2048MB for working memory.  All other parameters were
left at their default values.  The computer used in the experiments
has an Intel Xeon E5440 2.83GHz processor.

The computational experiments compare the \ooyprime\ model with our
proposed model.  In a set of prelimary experiments, CPLEX was not able
to find a feasible solution to a few of the instances within the time
limit, regardless of the model used.  To remedy this, we gave CPLEX an
initial feasible solution produced by a constructive heuristic we call
EST, for ``earliest starting time''.  (The makespan of the schedule
found by EST was also used to set the value of parameter~$L$.)

\begin{table}[bt]
  \centering
{\footnotesize
  \begin{tabular}[ht]{|l|c|c|c|c|c|c|}
    \hline
    \multirow{2}{*}{Instance} & \multirow{2}{*}{Size} &
    \multirow{2}{*}{EST} &
    \multicolumn{2}{c}{\rule{0ex}{2.5ex}\ooyprime\ model} & 
    \multicolumn{2}{|c|}{new model} \\
    \cline{4-7}
    & & & $\mks$ & CPU(s) & $\mks$ & CPU(s)\\
    \hline
    SFJS01 & 2, 2, 2 & 66 & 66 & 0.01 & 66 &  \graycell 0.00\\
    SFJS02 & 2, 2, 2 & 107 & 107 & 0.01 & 107 &  0.01\\
    SFJS03 & 3, 2, 2 & 255 & 221 &  \graycell 0.02 & 221 & 0.05\\
    SFJS04 & 3, 2, 2 & 367 & 355 & 0.02 & 355 &  0.02\\
    SFJS05 & 3, 2, 2 & 143 & 119 & 0.04 & 119 &  0.04\\
    SFJS06 & 3, 3, 2 & 360 & 320 & 0.03 & 320 & \graycell 0.01\\
    SFJS07 & 3, 3, 5 & 407 & 397 & 0.01 & 397 & \graycell 0.00\\
    SFJS08 & 3, 3, 4 & 273 & 253 & \graycell 0.03 & 253 & 0.04\\
    SFJS09 & 3, 3, 3 & 230 & 210 &  0.02 & 210 & \graycell 0.01\\
    SFJS10 & 4, 3, 5 & 608 & 516 & \graycell 0.01 & 516 & 0.02\\
    MFJS01 & 5, 3, 6 & 526 & 468 & 0.53 & 468 &  \graycell 0.26\\
    MFJS02 & 5, 3, 7 & 540 & 446 & 1.20 & 446 &  \graycell 0.87\\
    MFJS03 & 6, 3, 7 & 655 & 466 & 4.09 & 466 &  \graycell 1.66\\
    MFJS04 & 7, 3, 7 & 690 & 554 & 126  & 554 &  \graycell 27.43\\
    MFJS05 & 7, 3, 7 & 690 & 514 & 5.70 & 514 &  \graycell 4.55\\
    MFJS06 & 8, 3, 7 & 838 & 634 & 1739 & 634 &  \graycell 52.48\\
    MFJS07 & 8, 4, 7 & 1130 & [859;881] 2.50\% & 3600 & 
    \graycell 879 &  \graycell 1890\\
    MFJS08 & 9, 4, 8 & 1129 & [764;884] 13.57\% & 3600 & 
    \graycell [775;884] 12.33\% & 3600\\
    MFJS09 & 11, 4, 8 & 1343 & \graycell [845.26;1104] 23.44\% & 3600
    &  [809.03;1137] 28.84\% & 3600\\
    MFJS10 & 12, 4, 8 & 1559 & [951.30;1263] 24.68\% & 3600
    & \graycell [944.80;1251] 24.48\% & 3600 \\ \hline
  \end{tabular}}
  \caption{\label{tab:fattahiH}
  Solving the Fattahi \etal~\cite{fattahi:07} instances
  with the \ooyprime\ and the new model.
  Both MILPs were solved using CPLEX,
  with initial feasible solution provided by the EST heuristic.
  The size of an instance is a triple $(n,o,m)$,
  where $n$~is the number of jobs, 
  $o$~is the the number of operations per job,
  and $m$~is the number of machines.
  All operations have integer processing times.
  The smallest CPU times and gaps are highlighted.}
\end{table}

\begin{table}[bt]
  \centering
{\footnotesize
  \begin{tabular}[ht]{|l|c|c|c|c|c|c|}
    \hline
    \multirow{2}{*}{Instance} & \multirow{2}{*}{Size} & 
    \multirow{2}{*}{EST} & 
    \multicolumn{2}{c}{\rule{0ex}{2.5ex}\ooyprime\ model} & 
    \multicolumn{2}{|c|}{new model} \\
    \cline{4-7}
    & & & $\mks$ & CPU(s) & $\mks$ & CPU(s)\\
    \hline
    MK01 & 10, 5-6, 6 & 49 & \graycell 40 & 1621 &
    [39;40] 2.5\% & 3600\\
    MK02 & 10, 5-6, 6 & 41 & [23;30] 23.33\% & 3600 &
     \graycell [25;29] 13.79\% & 3600\\
    MK03 & 15, 10, 8 & 204 & [63;204] 69.12\% & 3600 &
     \graycell [92;204] 54.90\% & 3600\\
    MK04 & 15, 3-9, 8 & 73 & [38;67] 43.28\% & 3600 &
     \graycell [41.09;65] 36.78\% & 3600\\
    MK05 & 15, 5-9, 4 & 186 &[58.85;186] 68.36\% & 3600 &
     \graycell [66.2;184] 64.02\% & 3600\\
    MK06 & 10, 15, 15 & 98 & [33;98] 66.33\% & 3600 &
     \graycell [37;98] 62.24\% & 3600\\
    MK07 & 20, 5, 5 & 214 &\graycell [62;174] 64.37\%  & 3600 &
    [61.17;192] 68.14\% & 3600\\
    MK08 & 20, 10-14, 10 & 523 &\graycell [181.25;523] 65.34\% & 3600 &
     [181;523] 65.39\% & 3600\\
    MK09 & 20, 10-14, 10 & 336 & [140.32;336] 58.24\% & 3600 &
    \graycell [146;336] 56.55\% & 3600\\
    MK10 & 20, 10-14, 15 & 274 & [104;274] 62.04\% &  3600 &
    \graycell [119;274] 56.57\% & 3600\\
    MK11 & 30, 5-7, 5 & 698 & \graycell [158.88;695] 77.14\% & 3600 &
     [152;698] 78.22\% & 3600\\
    MK12 & 30, 5-9, 10 & 566 & [160;524] 69.47\% & 3600 &
    \graycell [180;546] 67.03\% & 3600\\
    MK13 & 30, 5-9, 10 & 500 & \graycell [153;482] 68.26\% &  3600 &
    [157;500] 68.60\% & 3600 \\
    MK14 & 30, 8-11, 15 & 719 & [228.97;719] 68.15\% & 3600 &
    \graycell [232;719] 67.73\% & 3600\\
    MK15 & 30, 8-11, 15 & 443 & [154;443] 65.24\% &  3600 &
    \graycell [190;443] 57.11\% & 3600\\ \hline
  \end{tabular}}
  \caption{Solving the Brandimarte~\cite{brandimarte:93} instances
  through the \ooyprime\ and the new model.
  Both MILPs were solved by CPLEX,
  with initial feasible solution provided by the EST heuristic.
  All operations have integer processing times.
  The smallest gaps are highlighted.
  }
\label{tab:brandimarteH}
\end{table}

\begin{table}[bt]
  \centering
{\footnotesize
  \begin{tabular}[ht]{|l|c|c|c|c|c|c|}
    \hline
    \multirow{2}{*}{Instance} & \multirow{2}{*}{Size} & 
    \multirow{2}{*}{EST} & 
    \multicolumn{2}{c}{\rule{0ex}{2.5ex}\ooyprime\ model} & 
    \multicolumn{2}{|c|}{new model} \\
    \cline{4-7}
    & & & $\mks$ & CPU(s) & $\mks$ & CPU(s)\\
    \hline
    YFJS01 & 4, 10, 7 & 1318 & 773 & 68.37 & 
    773 &  \graycell 11.5\\
    YFJS02 & 4, 10, 7 & 1243 & 825 & \graycell 6.03 & 
    825 &  9.88\\
    YFJS03 & 6, 4, 7 & 439& 347 & 11.11 &
    347 & \graycell 3.72\\
    YFJS04 & 7, 4, 7 & 569 & 390 & 22.63 &
    390 & \graycell 7.82\\
    YFJS05 & 8, 4, 7 & 566 & 445 & 660.79 &
    445 & \graycell 357.55\\
    YFJS06 & 9, 4, 7 & 633 & [378.90;452] 16.17\% & 3600 &
    \graycell [425.29;449] 5.28\% & 3600\\
    YFJS07 & 9, 4, 7 & 628 & [439;460] 4.57\% & 3600 & 
    \graycell 444 & \graycell 1392\\
    YFJS08 & 9, 4, 12 & 531 & 353 & 3.26 & 
    353 & \graycell 0.67\\
    YFJS09 & 9, 4, 12 & 506 & [238;242] 1.65\% & 3600 &
    \graycell 242 & \graycell 14.03\\
    YFJS10 & 10, 4, 12 & 541 & 399 & 22.21 &
    399 & \graycell 4.03\\
    YFJS11 & 10, 5, 10 & 740 & 526 & 1699.64 &
     526 & \graycell 177.43\\
    YFJS12 & 10, 5, 10 & 813 & [465;541] 14.05\% & 3600 &
    \graycell 512 & \graycell  3218.89\\
    YFJS13 & 10, 5, 10 & 717 & [376;405] 7.16\%  & 3600 &
    \graycell 405 & \graycell 1624.66\\
    YFJS14 & 13, 17, 26 & 2055 & [1317;1461] 9.86\% & 3600 &
    \graycell 1317 & \graycell 3293.58\\
    YFJS15 & 13, 17, 26 & 2296 & [1239;1260] 1.67\% & 3600 &
    \graycell [1239;1244] 0.40\% & 3600\\
    YFJS16 & 13, 17, 26 & 2006 & [1189;1432] 16.97\% & 3600 &
    \graycell [1200;1245] 3.61\% & 3600\\
    YFJS17 & 17, 17, 26 & 2408 & \graycell [1133;1832] 38.16\% & 3600 &  
    [1133;2379] 52.37\% & 3600\\
    YFJS18 & 17, 17, 26 & 2082 & \graycell [1220;1772] 31.15\% & 3600 &  
    [1220;2082] 41.40\% & 3600\\
    YFJS19 & 17, 17, 26 & 2038 & [862;1897] 54.56\% & 3600 & 
    \graycell [926;1581] 41.43\% & 3600\\
    YFJS20 & 17, 17, 26 & 2369 & [705;1686] 58.19\% & 3600 & 
    \graycell [968;2312] 58.13\% & 3600\\
    \hline
  \end{tabular}}
  \caption{Solving FJS instances that consist of Y-jobs
  with the \ooyprime\ and the new model.
  Both MILPs were solved by CPLEX,
  with initial feasible solution provided by the EST heuristic.
  All makespans are integer since all processing times are integer.
  (Curiously, many of the lower bounds are also integer.)
  The smallest CPU times and gaps are highlighted.
  }
  \label{tab:newinstancesH}
\end{table}


The EST heuristic produces a permutation
$(v_1,v_2,\ldots,v_{\NN})$ of the set $V$ of operations 
such that $i<j$ whenever $(v_i,v_j) \in A$, 
and a corresponding sequence $(\ff_1,\ff_2,\ldots,\ff_{\NN})$ of machines.  
Of course $\ff_i\in F(v_i)$ for each~$i$, 
and $v_i$ is to be processed on machine~$\ff_i$.  
This pair of sequences defines an admissible selection: 
just take the set of all ordered pairs $(v_i,v_j)$ in $V \times V$ 
such that $i<j$ and $\ff_i=\ff_j$.

Each iteration of EST starts with sequences $(v_1,\ldots,v_{q-1})$
and $(\ff_1,\ldots,\ff_{q-1})$ such that no arc in $A$ enters the 
set $U := \conj{v_1,\ldots,v_{q-1}}$. Let $A_U := A\cap(U\times U)$,
and let $Y_U$ be the set of all pairs 
$(v_i,v_j)$ in $U\times U$
such that $i<j$ and $\ff_i=\ff_j$. 
Let $\sstar$ be the tight schedule for 
$(U, A_U\cup Y_U, \pprime)$, 
where $\pprime_{v_i} := \pp_{v_i,\ff_i}$.
Then the completion time of $v_i$ will be 
$\cc_i := \sstar_{v_i} + \pprime_{v_i}$ 
and each machine $\kk$ will become available at time
\[
\max\,\conj{\cc_i : 1\leq i < q \logicand \ff_i = \kk}
\]
(or $0$ if there is no $i$ such that $\ff_i = \kk$). 
From this information,
the heuristic chooses $v_{q}$ in $V\setminus U$ 
and $\ff_{q}$ in~$F(v_{q})$.
The idea is to choose a pair $(v_{q},\ff_{q})$
whose execution can start the earliest.
This rule is, in general, satisfied by several pairs.
Experience shows that an additional tie-breaking rule 
can significantly improve the heuristic.
Let $\meanpp$ be the mean processing time, 
i.e.\ $\meanpp_v := ({\sum_{\kk\in F(v)}\pp_{v,\kk}})/{\abs{F(v)}}$ 
for each operation~$v$. 
The tie-breaking rule can be stated as follows.
If there are several candidate pairs $(w,\kk)$ 
for the role of $(v_{q},\ff_{q})$,
choose a pair that maximizes
the largest sum of the form
$\meanpp_w + \meanpp_{z_1}+\meanpp_{z_2}+\cdots+\meanpp_{z_{\myel}}$,
where $(w,z_1,z_2,\ldots,z_{\myel})$ is a 
path in~$(V,A)$.
This heuristic takes time $\Oh(\abs{V}\abs{A} + \abs{V}^2\abs{M})$.


Table~\ref{tab:fattahiH} shows the results for 
the instances of Fattahi \etal~\cite{fattahi:07} 
and Table~\ref{tab:brandimarteH} 
for the instances of Brandimarte~\cite{brandimarte:93}. 
These two sets of instances contain
only path-jobs. Tables~\ref{tab:newinstancesH} and~\ref{tab:newinstancesG}
show the results for the two new sets of instances, 
generated as described above.  
The Instance column records the names 
of the instances.  
The Size column contains the number of jobs, 
the number of operations per job 
(or an interval $a$-$b$, when the jobs 
have between $a$ and $b$ operations), 
and the number of machines. 
Column EST shows the makespan of the schedule produced by the heuristic. 
For each model, the $\mks$ column records 
the optimal
makespan or the lower and upper bounds 
found by CPLEX, 
followed by the relative gap (the difference between upper and lower
bounds, divided by the upper bound). 
The CPU column records the CPU time, in seconds, 
taken by CPLEX to solve the MILP 
(or~3600, if CPLEX did not solve the MILP in one hour). 
In all cases, the EST heuristic 
used no more than a hundredth of a second.

In Table~\ref{tab:fattahiH} we observe that, for most instances, CPLEX
ran fastest under the new model.  Moreover, CPLEX found an optimal
solution for instance MFJS07 with the new model, while it did not find
one (within the time limit) with \ooyprime.  On the other hand, for
instance MFJS09, CPLEX obtained a solution with better makespan and
better lower bound using \ooyprime\ than using the new model.  Recall
that model \ooyprime\ reduces to the original \ooy\ model when applied
to the instances considered in Table~\ref{tab:fattahiH} and that,
moreover, numerical experiments of \Ozguven~\etal~\cite{ozguven:10}
show that their model is several orders of magnitude faster than the
model introduced in~Fattahi~\etal~\cite{fattahi:07}.

The instances in Table~\ref{tab:brandimarteH} are larger, in terms of
total number of operations, and clearly more difficult.  Under the new
model, CPLEX obtained better bounds for most of them.  For the
instance MK01, it found an optimal solution in just 60 seconds but
could not raise the lower bound to prove the optimality of the
solution.  Under the \ooyprime\ model, CPLEX found an optimal solution
in 1621 seconds.

\begin{table}[bt]
  \centering
{\footnotesize
\begin{tabular}[ht]{|l|c|c|c|c|c|c|}
    \hline
    \multirow{2}{*}{Instance} & \multirow{2}{*}{Size} & 
    \multirow{2}{*}{EST} & 
    \multicolumn{2}{c}{\rule{0ex}{2.5ex}\ooyprime\ model} & 
    \multicolumn{2}{|c|}{new model} \\
    \cline{4-7}
    & & & $\mks$ & CPU(s) & $\mks$ & CPU(s)\\
      \hline
      DAFJS01 & 4, 5-9, 5 & 327 & [255.0;257] 0.78\% & 3600 & \graycell 257 & \graycell 78.93\\
      DAFJS02 & 4, 5-7, 5 & 382 & [270.0;297] 9.09\% & 3600 & \graycell 289 & \graycell 1271.7\\
      DAFJS03 & 4, 10-17, 10 & 710 & 576 & 371.39 & 576 & \graycell 15.8\\
      DAFJS04 & 4, 9-14, 10 & 653 & 606 & 18.2 & 606 & \graycell 1.22\\
      DAFJS05 & 6, 5-13, 5 & 482 & \graycell [355.01;411] 13.62\% & 3600 & [347.53;403] 13.76\% & 3600\\
      DAFJS06 & 6, 5-13, 5 & 489 & [326;446] 26.91\% & 3600 & \graycell [326;435] 25.06\% & 3600\\
      DAFJS07 & 6, 7-23, 10 & 717 & [491.11;717] 31.5\% & 3600 & \graycell [497;562] 11.57\% & 3600\\
      DAFJS08 & 6, 6-23, 10 & 847 & [517;690] 25.07\% & 3600 & \graycell [628;631] 0.48\% & 3600\\
      DAFJS09 & 8, 4-9, 5 & 535 & [315;497] 36.62\% & 3600 & \graycell [315;475] 33.68\% & 3600\\
      DAFJS10 & 8, 4-11, 5 & 629 & \graycell [336;567] 40.74\% & 3600 & [336;575] 41.57\% & 3600\\
      DAFJS11 & 8, 10-23, 10 & 708 & [658;708] 7.06\% & 3600 & [658;708] 7.06\% & 3600\\
      DAFJS12 & 8, 9-22, 10 & 720 & [530;720] 26.39\% & 3600 & [530;720] 26.39\% & 3600\\
      DAFJS13 & 10, 5-11, 5 & 766 & [252;751] 66.44\% & 3600 & \graycell [304;718] 57.66\% & 3600\\
      DAFJS14 & 10, 4-10, 5 & 871 & [313;866] 63.86\% & 3600 & \graycell [358.95;860] 58.26\% & 3600\\
      DAFJS15 & 10, 8-19, 10 & 818 & [497;818] 39.24\% & 3600 & \graycell [512;818] 37.41\% & 3600\\
      DAFJS16 & 10, 6-20, 10 & 831 & [462;831] 44.4\% & 3600 & \graycell [640;819] 21.86\% & 3600\\
      DAFJS17 & 12, 4-11, 5 & 910 & [300;910] 67.03\% & 3600 & \graycell [300;909] 67.0\% & 3600\\
      DAFJS18 & 12, 5-9, 5 & 951 & [322;951] 66.14\% & 3600 & [322;951] 66.14\% & 3600\\
      DAFJS19 & 8, 7-13, 7 & 601 & [512;601] 14.81\% & 3600 & \graycell [512;592] 13.51\% & 3600\\
      DAFJS20 & 10, 6-17, 7 & 815 & [399;815] 51.04\% & 3600 & \graycell [434;815] 46.75\% & 3600\\
      DAFJS21 & 12, 5-16, 7 & 965 & [504;965] 47.77\% & 3600 & [504;965] 47.77\% & 3600\\
      DAFJS22 & 12, 5-17, 7 & 902 & [464;902] 48.56\% & 3600 & [464;902] 48.56\% & 3600\\
      DAFJS23 & 8, 6-17, 9 & 632 & [450;605] 25.62\% & 3600 & \graycell [450;538] 16.36\% & 3600\\
      DAFJS24 & 8, 6-25, 9 & 674 & [476;674] 29.38\% & 3600 & \graycell [476;666] 28.53\% & 3600\\
      DAFJS25 & 10, 9-19, 9 & 897 & [584;897] 34.89\% & 3600 & [584;897] 34.89\% & 3600\\
      DAFJS26 & 10, 8-17, 9 & 903 & [565;903] 37.43\% & 3600 & [565;903] 37.43\% & 3600\\
      DAFJS27 & 12, 7-22, 9 & 981 & [503;981] 48.73\% & 3600 & [503;981] 48.73\% & 3600\\
      DAFJS28 & 8, 8-15, 10 & 703 & [535;695] 23.02\% & 3600 & \graycell [535;671] 20.27\% & 3600\\
      DAFJS29 & 8, 7-19, 10 & 760 & [609;753] 19.12\% & 3600 & \graycell [609;726] 16.12\% & 3600\\
      DAFJS30 & 10, 8-19, 10 & 657 & [401;657] 38.96\% & 3600 & \graycell [467;656] 28.81\% & 3600\\
  \hline
  \end{tabular}}
  \caption{Solving the more general FJS instances through 
  the \ooyprime\ and the new model.
  Both MILPs were solved using CPLEX,
  with initial feasible solution provided by the EST heuristic.
  All makespans are integer since all processing times are integer.
  The smallest CPU times and gaps are highlighted.}
  \label{tab:newinstancesG}
  \end{table}

Table~\ref{tab:newinstancesH} shows that,
on instances with Y-jobs, 
CPLEX performed better with the new model,
running significantly faster and obtaining better bounds,
except on instances YFJS17 and YFJS18. 
Most of the instances in Table~\ref{tab:newinstancesG} 
seem to be more difficult.
CPLEX found an optimal solution for instances DAFJS01 and DAFJS02 with the
new model, 
while it could not find one (within the time limit) with \ooyprime. 
For instances DAFJS03 and DAFJS04, CPLEX found an optimal
solution under the new model considerably faster 
than it did under \ooyprime. 
For the other 26 instances, CPLEX did not find
an optimal solution with either model. For only two of these~26
instances, CPLEX achieved better bounds when using the \ooyprime\ model. For
eight of the~26 instances, CPLEX ended with the same bounds for both models,
and for~16 of the~26 instances, CPLEX achieved better bounds when using
the new model.

Here is a summary of our results.
For 34 of the 85 instances considered,
CPLEX found an optimal solution and proved its optimality
when using the new model.
Using the \ooyprime\ model, CPLEX did so for 27 of the 85 instances. 
For all but 7 of the 85 instances, 
CPLEX produced at least as good a lower bound
under the new model as under \ooyprime. 
For 27 of the 85 instances, CPLEX was strictly faster under the new model. 
For 11 of the 85 instances, CPLEX was faster under the \ooyprime\ model,
and for 49 instances, under both models, CPLEX ran for one hour
without finding an optimal solution. 
On the instances with Y-jobs and the 
more general ones, CPLEX performed especially well under the new model.
Taking all this into account, 
we conclude that, in general,
CPLEX produced better results with the new 
model than with~\ooyprime.

\section{Conclusion}\label{final}

The FJS problem is a generalization of the JS problem in which there may be
several machines, not necessarily identical, 
capable of processing an operation. 
In the literature on the problem, each job consists of 
a sequence of operations to be processed 
in a given order, as in the ordinary JS problem. 
In the present paper, 
we extended the definition of the FJS problem to allow 
an arbitrary precedence relation over the set of operations
and we presented a new MILP model for the extended problem. 
We also presented computational experiments indicating 
that the proposed model is better than that of \OzguvenOY~\cite{ozguven:10}. 
Some of our experiments were done on a new set of instances, 
inspired by a real application.
This set can be used as benchmark in future computational experiments
on the FJS problem.

For benchmarking purposes, and to allow reproduction 
of the results presented in this paper, 
the C/C++ code for the MILP models
(using the IBM ILOG CPLEX Concert Technology, version 12.1), 
as well as the code of the EST heuristic, 
the code of the two generators, 
and the four sets of instances used in the experiments
are available for download at
\url{http://www.ime.usp.br/~cris/fjs/}.

\smallskip
\noindent{\bf Acknowledgements.} 
We thank Jun Zeng for several suggestions and discussions. 

\newcommand{\etalchar}[1]{$^{#1}$}

\end{document}